\documentclass[aps,prl,twocolumn,showpacs,10pt,superscriptaddress,groupedaddress, footinbib]{revtex4-1}

\usepackage{amsmath,bm,amssymb}
\usepackage{amsbsy}
\usepackage{autobreak}
\usepackage[normalem]{ulem}
\usepackage{endnotes}
\let\footnote=\endnote

\usepackage{graphicx}
\usepackage{psfrag}
\usepackage[caption=false]{subfig}
\usepackage{color}
\definecolor{urlblue}{rgb}{0.2,0.4,0.7}
 \definecolor{citegreen}{rgb}{0,0.6,0.2}
\usepackage[caption=false]{subfig}
\usepackage{color}
\definecolor{urlblue}{rgb}{0.2,0.4,0.7}
\definecolor{citegreen}{rgb}{0,0.6,0.2}
\definecolor{linkred}{rgb}{0.9,0.2,0.1}
\usepackage{hyperref}
\hypersetup{
colorlinks=true, citecolor=citegreen, linkcolor=blue,urlcolor=urlblue}
\usepackage[table]{xcolor}
\usepackage{tabularx}
\newcolumntype{P}[1]{>{\centering\arraybackslash}p{#1}}
\usepackage{multirow, caption, booktabs}
\usepackage{slashed}
\usepackage{array}
\newcolumntype{P}[1]{>{\centering\arraybackslash}p{#1}}
\allowdisplaybreaks

\definecolor{linkred}{rgb}{0.9,0.2,0.1}

\usepackage{hyperref}
\hypersetup{
colorlinks=true, citecolor=citegreen, linkcolor=blue,urlcolor=urlblue}

\usepackage[table]{xcolor}
\usepackage{tabularx}
\newcolumntype{P}[1]{>{\centering\arraybackslash}p{#1}}
\usepackage{multirow, caption, booktabs}

\usepackage{slashed}
\usepackage{array}
\newcolumntype{P}[1]{>{\centering\arraybackslash}p{#1}}

\allowdisplaybreaks

\begin{document}

\preprint{IMSc/2017/08/07}

\title{Going beyond Soft plus Virtual}

\author{A.H. Ajjath$^{\scriptstyle 1}$}
\email{aabdulhameed@lpthe.jussieu.fr}
\author{Pooja Mukherjee,$^{\scriptstyle 2}$}
\email{pmukherj@uni-bonn.de}
\author{V. Ravindran,$^{\scriptstyle 3}$}
\email{ravindra@imsc.res.in}
\affiliation{$^{\scriptstyle 1}$Laboratoire de Physique Th\'eorique et Hautes Energies (LPTHE), UMR 7589, Sorbonne Universit\'e et CNRS, 4 place Jussieu, 75252 Paris Cedex 05, France}
\affiliation{$^{\scriptstyle 2}$Bethe Center for Theoretical Physics, Universit\"at Bonn, 53115 Bonn, Germany}
\affiliation{$^{\scriptstyle 3}$The Institute of Mathematical Sciences, HBNI, Taramani,
 Chennai 600113, India}

\date{\today}

\begin{abstract}

\textcolor{black}{We present a formalism that sums up the soft-virtual (SV) and next-to-SV (NSV) diagonal contributions to inclusive colorless productions in hadron colliders to all orders in perturbative QCD. Using factorisation theorem, renormalisation group invariance and employing the transcendental structure of perturbative results, we show the exponential behavior of soft-collinear function. This allows us to predict certain SV and NSV terms to all orders from lower order information. We also present an integral representation for the coefficient functions which is suitable for Mellin $N$-space resummation.}

\end{abstract}
\maketitle
 {\it Introduction: ---}The tests \cite{Brooijmans:2018xbu} of Standard Model (SM) of high energy physics has been going on at the Large Hadron Collider (LHC) with an unparalleled accuracy. 
Together, precise theoretical predictions of 
several of the observables   with strong and electroweak radiative corrections   are also available.  
Perturbative Quantum Chromodynamics (QCD) results  for both inclusive \cite{Anastasiou:2015ema,
Mistlberger:2018etf,Duhr:2019kwi,Duhr:2020seh} and differential observables to third order
in the strong coupling constant 
play important
role in precision studies. 
These perturbative results  improve
our understanding of ultraviolet and infrared structure of the underlying quantum field theory, \cite{Catani:1998bh,Becher:2009cu,Becher:2009qa,
Gardi:2009qi,Catani:1998bh,Ajjath:2019vmf,PhysRevD.100.114016}.  In particular,
the factorisation properties of amplitudes and cross sections and the corresponding renormalisation group (RG) equations  shed light on certain universal structures of the underlying dynamics which help us to sum up certain dominant contributions to all orders in perturbation theory
\cite{Moch:2005ky,Ravindran:2005vv,Ravindran:2006cg,deFlorian:2012za,Ahmed:2014cha, Kumar:2014uwa,Ahmed:2014cla,Catani:2014uta,Li:2014bfa}.
The factorisation of ultraviolet (UV) and infrared (IR) sensitive terms 
in the Green's function or in the observables bring in unphysical scales and their RGs 
are controlled by the universal anomalous dimensions. 
In the seminal works by Sterman  \cite{Sterman:1986aj} and by Catani and Trentadue \cite{Catani:1989ne}
the contributions from large logarithms from soft gluons were shown to exponentiate in a systematic fashion.  Remarkable success  \cite{Bonvini:2016fgf,H:2019dcl,Moch:2005ba,Bonvini:2010ny,Bonvini:2012sh,H.:2020ecd}  in the resummation
of soft gluons lead to questions related to summing up of  subleading threshold logarithms, for example logarithms resulting from 
next to soft-virtual (NSV) contributions. 
In QCD and in soft collinear effective theory there have been significant developments to resum  NSV  terms to all orders
\cite{Laenen:2008ux,Laenen:2010uz,Bonocore:2014wua,Bonocore:2015esa,Bonocore:2016awd,DelDuca:2017twk,Bahjat-Abbas:2019fqa,Soar:2009yh,Moch:2009hr,deFlorian:2014vta,Beneke:2018gvs,Bahjat-Abbas:2019fqa,Beneke:2019mua,Beneke:2019oqx}.
In this letter, restricting to the diagonal channels of the inclusive production of a colorless particle, we provide a framework to resum next to soft-virtual terms to all orders in perturbation
theory using
the mass factorisation, RG invariance and transcendentality structure of fixed order predictions.
{\color{black}We provide an elaborate discussion on the structure of NSV logarithms and  on the resummation formalism in the longer version \cite{Ajjath:2020ulr}.}

\textcolor{black}{{\it Factorisation: ---} We consider the inclusive cross sections for the production of color-singlet final states,  such as the production of a single scalar Higgs boson in gluon fusion or in bottom quark annihilation and lepton pair production in Drell-Yan (DY) process.  
In the QCD improved parton model, thanks to the well established factorization theorem for the inclusive cross sections, the hadronic cross section $\sigma(q^2,\tau)$ can be
expressed in terms of  mass factorised partonic  coefficient functions (CFs), $\Delta_{ab}(q^2,\mu_R^2,\mu_F^2,z)$ and 
parton distribution functions (PDFs), $f_c(x_i,\mu_F^2)$, of incoming partons:  
\begin{align}
\label{QCDpm}
\sigma(q^2,\tau) = \sigma_0(\mu_R^2) \sum_{ab}& \int dx_1 \int dx_2 f_a(x_1,\mu_F^2) f_b(x_2,\mu_F^2) \nonumber \\
&\times \Delta_{ab}(q^2,\mu_R^2,\mu_F^2,z) \,,
\end{align}
with $\sigma_0$ being the born level cross section.
The hadronic scaling variable is defined by $\tau=q^2/S$, where $S$ is square of the hadronic center of mass energy.   Similar scaling variable of CF at the partonic level is denoted by $z=q^2/\hat{s}$, with $\hat{s}$ being the partonic center of mass energy. . Here $q^2$ refers to the invariant mass of 
final state leptons, $M^2_{l^+l^-}$, for DY and for the Higgs boson productions  $q^2=m_H^2$, with $m_H$ being the Higgs boson mass.
The subscripts $a,b$ in $\Delta_{ab}$
and $c$ in $f_c$ collectively denote the type of parton (quark, antiquark and gluon), 
their flavour etc.   The hadronic and partonic scaling variables are related through $\hat{s}=x_1 x_2 S$, which in turn implies $z=\tau/(x_1 x_2)$ with $x_i$ refers to
the  momentum fraction of the incoming partons.}

\textcolor{black}{The inclusive cross sections beyond leading order in perturbation theory
contain collinear singularities resulting from massless initial states.  The mass factorisation theorem allows one to decompose  such cross sections in terms of collinear singular but universal/process independent  Altarelli-Parisi (AP)
kernels \cite{Altarelli:1977zs}, $\Gamma_{ab}$, and process dependent finite CFs, $\Delta_{ab}$, 
at an arbitrary factorisation scale $\mu_F$:}
\begin{align}
\label{MassFact}
{1 \over z} {\hat \sigma_{ab} (q^2,z,\epsilon) } =& \sigma_0(\mu_R^2)\sum_{a' b'}  \Gamma^T_{a a'}(z,\mu_F^2,) 
\nonumber\\
& \otimes
{\Delta_{a'b'}(q^2,\mu_R^2,\mu_F^2,z,\epsilon)} \otimes \Gamma_{b'b}(z,\mu_F^2,\epsilon) \,. 
\end{align}

\textcolor{black}{The PDFs given in \eqref{QCDpm} are related to bare PDFs $\hat f_b$ by AP kernels, i.e.,
$f_a(\mu_F^2)=\Gamma_{ab}(\mu_F^2)\otimes \hat f_b$.  The CFs are expanded in powers of coupling constant
$a_s(\mu_R^2)=g_s^2(\mu_R^2)/16 \pi^2$ as $\Delta_{ab}=\sum_{i} a_s^i(\mu_R^2) \Delta_{ab}^{(i)}(\mu_R^2)$. The $g_s$ is
renormalised strong coupling constant of QCD and $\mu_R$ is the renormalisation scale}

 \textcolor{black}{The CFs, $\Delta_{ab}$, can be classified into two categories, viz, diagonal (${\rm CF}_d$) when $b=\overline a$ and off-diagonal (${\rm CF}_{nd}$).  These CFs depend on two unphysical
scales $\mu_F,\mu_R$, a physical scale $q^2$ and the scaling variable $z$.  
In the following, we will investigate
the all order perturbative structure of CFs in terms of $q^2$ and the scaling variable $z$ by setting
up a Sudakov type of differential equation for CFs in the kinematic region where $z$ is closer to threshold limit $z=1$.   Let us  begin with the mass factorisation for ${\rm CF}_d$s, for example, 
of the DY process:  }
\begin{eqnarray}\label{massfactDY}
   \frac{\hat\sigma_{q \bar q}}{z\sigma_0}= & ~
    \Gamma_{qq}^T \otimes  \Delta_{q\bar q}
 \otimes \Gamma_{\bar q \bar q}+
  \Gamma_{qq}^T \otimes \Delta_{qg}
 \otimes \Gamma_{g \bar q} 
+ \cdots \,.
 \end{eqnarray}
\textcolor{black}{If we restrict only to distributions such as ${\cal D}_k(z)=(\ln^k(1-z)/(1-z))_+, ~
k \ge 0$,
and $\delta(1-z)$, the SV terms and   $L^k_z=\ln^k(1-z), \text{ with } k\ge 0$,called  NSV terms, then  only the first term
in the above expansion survives. 
Rest of the terms in \eqref{massfactDY} contains at least one pair of non-diagonal pieces which upon convolutions
will give terms of the form $(1-z)^l \ln^k(1-z) ,l>0,k\ge 0$.  They are called beyond NSV contributions and are not considered in our study.}

\textcolor{black}{Further, the diagonal $\Gamma_{q q}$s in the first term in \eqref{massfactDY}
also contain beyond NSV terms, dropping the later will give rise to
a simple form for the $\Gamma_{q q}$s containing only diagonal AP splitting functions, $P_{cc}$.
This is true for $\hat \sigma_{b \overline b}$ and $\hat \sigma_{gg}$
in the threshold limit.
In summary, for diagonal channels, the mass factorised result given in \eqref{MassFact} contains only diagonal 
terms $\hat \sigma_{c \overline c}$, $\Delta_{c \overline c}$ and AP kernels $\Gamma_{cc}$ and 
the sum over $ab$ is dropped:}
\begin{eqnarray}
{\hat \sigma_{c \bar c}^{\mathrm{sv+nsv}}  \over z\sigma_0} =\Gamma_{c c}^T\otimes \Delta_{c \bar c}^{\mathrm{sv+nsv}} \otimes \Gamma_{\bar c \bar c}\,.  
\end{eqnarray}
\textcolor{black}{We will show below that this remarkable simplification happens only for the diagonal CFs, allowing us to
explore their perturbative structure with the help of Sudakov $K+G$ type of first order differential equation with respect to $q^2$.  }

\textcolor{black}{For an off-diagonal  channel, say $\hat \sigma_{qg}$, we find}
\begin{align}\label{massfactqg}
   \frac{\hat\sigma_{q g}}{z\sigma_0} = &~
    \Gamma_{qq}^T \otimes \Delta_{qq}
 \otimes \Gamma_{q g}
 +
  \Gamma_{qq}^T \otimes \Delta_{qg}
 \otimes \Gamma_{g g} + \cdots \,.
\end{align} 
\textcolor{black}{In the above expansion, no single term produces distributions after the convolution, since each term contains at least one off-diagonal term.   If we then restrict to NSV contributions, those involving
at least two off-diagonal pieces do not contribute, hence results in :}
\begin{align}
\frac{\hat \sigma_{qg}^{\mathrm{sv+nsv}}}{z\sigma_0} =\Gamma_{q q}^T \otimes \Delta_{q \overline q}^{\mathrm{sv+nsv}}\otimes \Gamma_{\overline q g}  + \Gamma_{q q}^T\otimes\Delta_{qg}^{\mathrm{sv+nsv}} \otimes\Gamma_{gg} \,.
\end{align}
Note that the off-diagonal $\Delta_{qg}$ receives contributions from $\hat \sigma_{qg}$ as well as from $\Delta_{q \overline q}$ unlike the diagonal $\Delta_{q \overline q}$ which receives only from $\hat \sigma_{q \overline q}$.  This feature makes the diagonal ones simpler than the rest.  The rest of the article will only deal with ${\rm CF}_d$s, unless otherwise stated.

\textcolor{black}{{\it Coefficient function: ---}
The ${\rm CF}_d$s of inclusive cross sections get contributions from form factor (FF) type processes, where the final state contains only
colorless particle(s), and from those processes {\color{black}which involve at least one real parton emission}.  The former from the FF, such as
$\hat F_c, c=q,g,b$ 
is proportional to $\delta(1-z)$ and hence can be factored out from $\hat \sigma^{\rm sv+nsv}_{c \overline c}$ along with the square of UV renormalisation constant $Z_{UV,c}$,if any. We call the resulting one by Soft-Collinear function, that is:}
\begin{widetext}
\begin{small}
\begin{eqnarray}
\label{normSc}
{\cal S}_{c}(\hat a_s,\mu^2,q^2,z,\epsilon) &=& \left(\sigma_0(\mu_R^2) \right)^{-1}
	\left(Z_{UV,c}(\hat{a}_s,\mu_R^2,\mu^2,\epsilon) \right)^{-2}
|\hat F_{c}(\hat a_s,\mu^2,-q^2,\epsilon)|^{-2}  \delta(1-z) \otimes \hat \sigma^{\rm sv+nsv}_{c \overline c} (q^2,z,\epsilon) 
\end{eqnarray}
\end{small}
\end{widetext}
\textcolor{black}{Note that the function ${\cal S}_{c}$
is computable in perturbation theory in powers of $\hat a_s$ and is  RG invariant with respect to $\mu_R$.
Substituting  for $\hat{\sigma}_{c\bar{c}}$ from \eqref{normSc} in terms of ${\cal S}^c$, in \eqref{MassFact}  and keeping only the diagonal terms in AP kernels, we obtain $\Delta^{\rm{sv+nsv}}_{c \overline c} \equiv \Delta_c$:}
\begin{widetext}
\begin{eqnarray}
\label{MasterF}
\Delta_c(q^2,\mu_R^2,\mu_F^2,z) &=&  
\left(Z_{UV,c}(\hat{a}_s,\mu_R^2,\mu^2,\epsilon) \right)^2
|\hat F_{c}(\hat a_s,\mu^2,-q^2,\epsilon)|^2
\delta(1-z) 
\nonumber\\  &&
\otimes    \big(\Gamma^T\big)_{cc}^{-1}(z,\mu_F^2,\epsilon)\otimes\mathcal{S}_{c} (\hat a_s,\mu^2,q^2,z,\epsilon)\otimes
    \Gamma^{-1}_{\bar{c}\bar{c}}(z,\mu_F^2,\epsilon)\,.
\end{eqnarray}
\end{widetext}
\textcolor{black}{So far, we have shown that if we restrict ourselves to SV+NSV terms in the partonic CFs, the diagonal
CFs take simpler form compared to non-diagonal ones.  For the diagonal ones, ${\rm CF}_d$s decompose into building blocks such as
squares of FF and of UV {\color{black} renormalisation constant}, soft-collinear function and diagonal AP kernels. } 

\textcolor{black} {There is a great deal of understanding of the infrared and UV structure of the FFs through Sudakov K+G equation \cite{Sudakov:1954sw,Sen:1981sd,Collins:1989bt,Magnea:1990zb,Magnea:2000ss,Sterman:2002qn,Moch:2005id,Ravindran:2005vv} and of the AP kernels through AP evolution equation in terms of universal anomalous dimensions.
For the FF, the factorisation of IR singularity implies that $\hat F_c(q^2)
= Z_{\hat F_c}(q^2,\mu_s^2) F_{c,fin}(q^2,\mu_s^2)$, where $Z_{\hat F_c}$ is IR singular and $F_{c,fin}$ is IR finite, the scale $\mu_s$ is IR factorisation scale.    Differentiation w.r.t to $q^2$ leads to Sudakov K+G differential equation, namely $d\ln \hat F_c/d\ln(q^2) = (K_c+G_c)/2$, where the IR singular kernel $K_c(\mu_s^2) = d \ln Z_{\hat F_c}/d\ln(q^2)$ and IR finite $G_c(q^2,\mu_s^2) =d \ln F_{c,fin}/d\ln(q^2)$.  The solution to K+G equation:
}
\begin{eqnarray}
\label{Fcsol}
\hat F_c(-q^2,\epsilon) = \exp\left(\int_0^{-q^2} {d \lambda^2 \over \lambda^2} \Gamma_{\hat F,c}(\lambda^2,\epsilon)\right)
\end{eqnarray}
\textcolor{black}{
with $\hat F_c(-q^2=0,\epsilon)=1$ and $\Gamma_{\hat F,c} = (K_c+G_c)/2$ is the kernel.  The UV renormalisation
constant $Z_{UV,c}$ admits similar exponential solution governed by anomalous dimension $\gamma_{UV,c}$.  The later are known to third order in QCD for $c=b$, see \cite{Vermaseren:1997fq} and
for $c=g $,  see
\cite{Chetyrkin:2005ia}. 
The AP kernel $\Gamma_{cc}$ satisfies AP evolution equation and in the approximation we work with, they are
controlled only by diagonal AP slitting functions $P_{cc}$.  Hence the all order solution takes the simple form:}
\begin{eqnarray}
\label{Gcsol}
\Gamma_{cc}(\mu_F^2,z,\epsilon)={\cal C} \exp\left({1\over 2} \int_0^{\mu_F^2} 
{d \lambda^2 \over \lambda^2} P_{cc}(\lambda^2,z,\epsilon)\right)
\end{eqnarray}
\textcolor{black}{The symbol ${\cal C}$ is defined in \cite{Ravindran:2005vv}. The AP splitting function is known to third order in perturbation theory and the SV distributions
and NSV logarithms present in them are controlled by universal cusp and collinear anomalous dimensions.}

\textcolor{black}{{\it Soft-collinear function: ---} Our next task is  to unravel the factorisation properties of soft-collinear function by setting up a differential equation in dimensional regularisation.  Differentiating both sides of \eqref{MasterF} with respect to $q^2$ and using K+G equation for the FF,
we obtain}
\begin{eqnarray}
\label{calSc}
q^2{d{\cal S}_c(q^2,z) \over dq^2} = \Gamma_{{\cal S},c}(q^2,z)\otimes {\cal S}_c(q^2,z)
\end{eqnarray}
where 
\begin{eqnarray}
\label{Scker}
\Gamma_{{\cal S},c}\!\!&=&\!\!q^2{d\over dq^2}\Big({\cal C} \ln \Delta_c(q^2,\mu_R^2,\mu_F^2,z) ) \nonumber\\
&&-\ln|\hat F_c(-q^2)|^2 \delta(1-z)\Big)
\end{eqnarray}
\textcolor{black}{The fact that ${\cal S}_c$ and 
${\hat F}_c$ are RG invariant w.r.t  $\mu_R$ and $\mu_F$ implies that the derivative w.r.t $q^2$ of $\Delta_c$ in the first term in
\eqref{Scker}  has to be a function of only $q^2$ and $z$.   While the first term is finite, the second
term will be proportional to singular $K_c$ and finite $G_c$ of the kernel  $\Gamma_{\hat F,c}$.  This allows us to decompose
the kernel $\Gamma_{{\cal S},c}$   into a singular $\overline K_c$ and finite $\overline G_c$ pieces to all orders in perturbation theory and write \eqref{calSc} as
$d {\cal S}_c(q^2,z)/d \ln (q^2) ={\cal S}_c(q^2,z)\otimes (\overline K_c(\mu_s^2,z) +\overline G_c(q^2,\mu_s^2,z))/2$.   We find that  $\overline K_c$ can depend only on $\mu_s$ and 
 process independent anomalous dimension, $A^c$ as it is proportional to the $K_c$ of the FF.  However $\overline G_c(q^2,\mu_s^2,z)$ will
contain the process dependent parts.   Here, the scale $\mu_s$ is arbitrary scale. The fact that $\overline K_c + \overline G_c$ 
decomposition is valid to all orders in perturbation theory implies that the ${\cal S}_c$ is factorisable, i.e., we can  write ${\cal S}_c(q^2,z) = Z_c(q^2,\mu_s^2,z)\otimes {\cal S}_{c,{\rm fin}}(q^2,\mu_s^2,z)$, and identify the IR singular
$\overline K_c=d\ln Z_c/d\ln(q^2)$ and IR finite $\overline G_c=d\ln {\cal S}_{c,fin}/d\ln(q^2)$. ${\cal S}_{c,{\rm fin}}$ is IR finite. $Z_c$ is IR singular
and the fact that it depends on  $\overline K_c(\mu_s^2)$ implies that we can fix only the structure of $\ln(q^2)$ terms in $Z_c$.  However, the complete singular structure of $Z_c$ and its dependence on $\mu_s$ and $q^2$ can be obtained by solving the renormalisation group:}
\begin{eqnarray}
\mu_s^2 {d Z_c(\mu_s^2,q^2,z)\over d \mu_s^2}  &=&\gamma_{{\cal S},c} (\mu_s^2,q^2,z)\otimes Z_c(\mu_s^2,q^2,z)
\end{eqnarray}
\textcolor{black}{where $\gamma_{{\cal S},c}$  takes the remarkable structure $\xi_1(\mu_s^2,z) \ln(q^2/\mu_s^2) + 
\xi_2(\mu_s^2,z)$ to all orders in perturbation theory.  This structure follows from the fact that $Z_c$ has to contain right infrared poles to cancel against those from FF and AP kernels leaving $\Delta_c$ finite.   We find that
}  
\begin{eqnarray}
\label{scanom}
\gamma_{{\cal S},c}&=&\Big(A^c(\mu_s^2) \ln\Big({q^2\over \mu_s^2}\Big)-{f^c(\mu_s^2)\over 2}\Big)\delta(1-z)+P'_{cc}(\mu_s)\nonumber\\
{\rm where}\nonumber\\
P'_{cc}&=&{2 A^c(\mu_s^2)\over (1-z)_+}
+2 C^c(\mu_s^2)\ln(1-z) + 2 D^c(\mu_s^2).
\end{eqnarray}
\textcolor{black}{Here $A^c,(D^c,C^c)$ and $f^c$ are cusp,collinear and soft anomalous dimensions respectively. 
The solution to ${\cal S}_c$ takes the form:}
\begin{eqnarray}
{\cal S}_c(q^2,z,\epsilon) &= &{\cal C} \exp\left(\int_0^{q^2} {d \lambda^2 \over \lambda^2} \Gamma_{{\cal S}_c}(\lambda^2,z,\epsilon)\right),
\nonumber\\
&=&{\cal C} \exp\left(2 \Phi_c(q^2,z,\epsilon)\right)
\end{eqnarray}
\textcolor{black}{
with ${\cal S}_c(q^2=0,z,\epsilon)=\delta(1-z)$.}

\textcolor{black}{{\it Transcendentality principle: ---} \textcolor{black}{
In the following, we study the logarithmic structure of $\Phi_c$ using the available fixed order results and propose an all order generalisation based on their remarkable transcendentality structure.
The  $\Gamma_{{\cal S},c}$ can be determined  
using $\Delta_c $ and $\hat F_c$ which are known to third order in $a_s$ and to desired accuracy in $\epsilon$ for  DY ($c=q$), Higgs boson production in gluon fusion ($c=g$) and in bottom quark annihilation ($c=b$), 
see  \cite{vanNeerven:1985xr,Harlander:2000mg,Ravindran:2004mb,Moch:2005tm,Gehrmann:2005pd,Baikov:2009bg,Gehrmann:2010ue,Gehrmann:2014vha,vonManteuffel:2016xki,Henn:2016men,Henn:2019rmi,vonManteuffel:2020vjv,Gehrmann:2010tu} and \cite{Anastasiou:2015ema,
Mistlberger:2018etf,Duhr:2019kwi,Duhr:2020seh}. In $\Delta_c$s, the explicit results to third order in $a_s$ show certain universal structure for  leading SV distribution as well as NSV logarithm, for example, at order $a_s^i$, both of them have
degree $2i$ independent of $c$.  Similarly, in FFs, computed in dimensional regularisation, 
if we assign $n_\epsilon$ weight for $\epsilon^{-n_\epsilon}$
and $n_\zeta$ for $\ln^{n_\zeta}(1-z)$, then the highest weight at every order
in $\epsilon$ shows uniform transcendentality
$\omega=n_\epsilon+n_\zeta$.
Hence, the explicit results for 
$\Gamma_{{\cal S}_c}$ obtained from $\Delta_c$ and $\hat F_c$ in dimensional regularisation also reveal the rich  structure for SV distributions and NSV logarithms  through transcendental weight.}}

\textcolor{black}{
We now turn to ${\cal S}_c$. Note that ${\cal S}_c$ is UV finite and hence a simple dimensional analysis implies that 
the $\Gamma_{{\cal S},c}$ can be  expanded
in powers of $\hat a_s (q^2/\mu^2)^{\epsilon/2}$.  The fact that $\Delta_c$ is finite implies  the soft-collinear function ${\cal S}_c$ has to contain right soft and collinear singularities
to cancel against those from FF and the AP kernels. These singularities appear as poles in $\epsilon$  resulting from the Feynman loop and phase space integrals.  In \cite{Ravindran:2004mb,Ravindran:2005vv}, the all order structure of SV part of
$\Gamma_{{\cal S},c}$ or equivalently SV part of $\Phi_c$ was determined. Here, we generalise this to include
NSV part by modifying $\Gamma_{{\cal S},c}$ in such a way that it contains additional collinear sensitive terms that
cancel collinear singularities from NSV part of AP kernels giving rise to right  NSV part of $\Delta_c$. Keeping RG invariance intact, we write }
\begin{eqnarray}\label{PhiSV1}
\mathrm{\Phi}^{c}(\hat{a}_s, q^2,\mu^2,z, \epsilon) &&=\mathrm{\Phi}^{c}_A + \mathrm{\Phi}^{c}_B
\nonumber\\&&
=\sum_{i=1}^\infty\hat{a}_s^i\Big(\frac{q^2(1-z)^2}{\mu^2 }\Big)^{i\frac{\epsilon}{2}} S_\epsilon^{i}
\Big(\frac{i\epsilon}{1-z}\Big)
\nonumber\\
&& \times \left(\hat{\phi}_{c}^{SV,(i)}(\epsilon)
+(1-z) \hat{\varphi}_{c}^{(i)}(z,\epsilon)\right).
\end{eqnarray}
\textcolor{black}{where $ S_\epsilon =\exp (\frac{\epsilon}{2}[\gamma_E - \ln (4\pi )])$
with $ \gamma_E $ being the Euler Mascheroni constant.  
The term $q^2 (1-z)^2$ inside the parenthesis is the scale corresponding to soft gluon emissions. Note that we have normalised the second term by this soft scale. The functions $\hat \phi^{SV,(i)}_c(\epsilon)$ and $\hat \varphi^{(i)}_c(z,\epsilon)$
contain poles in $\epsilon$.     
The first term, $\mathrm{\Phi}^{c}_A$ containing $(1-z)^{i \epsilon}/(1-z) \hat \phi^{SV,(i)}_c(\epsilon)$ is sufficient to obtain the right distributions ${\cal D}_j$ and $\delta(1-z)$ in $\Delta_c$,  and they constitute to the SV contributions to CF (see  \cite{Ravindran:2005vv,Ravindran:2006cg}). The 
NSV terms $\ln^k(1-z), k=0,\cdots$ in $\Delta_c$, on the other hand, are generated from first as well as the second term $\mathrm{\Phi}^{c}_B$ containing
$(1-z)^{i \epsilon}\hat \varphi^{(i)}_c(z,\epsilon)$.
Note that in $\mathrm{\Phi}^c_A$, the entire $z$ dependence
factors out leaving only $\hat \phi^{SV,(i)}_c(\epsilon)$
at every order.  This happens because the soft gluons factorise
at a single scale, namely $q^2 (1-z)^2$ at every order in $a_s$. Consequently, the entire series containing soft gluon contributions can be summed up to obtain exponential solution $\exp(2 \Phi^c_A)$.
Explicit computation of 
the exponent $\Phi^c_B$ demonstrates a peculiar dependence on the scaling variable $z$ through $\hat \varphi^{(i)}_c(z,\epsilon)$ at every order in $\hat a_s$, given an accuracy in $\epsilon$.  
In $\mathrm{\Phi}^c_B$, we find that the highest power of $\ln(1-z)$ is controlled by the order of $a_s$ and the accuracy in $\epsilon$.  In particular,
if we assign $n_\epsilon$ weight for $\epsilon^{-n_\epsilon}$
and $n_L$ for $\ln^{n_L}(1-z)$, then the highest weight at every order
in $a_s$ shows uniform transcendentality
$\omega=n_\epsilon+n_L$. For example, at the order $a_s$, we find $\omega=1$ irrespective of the accuracy in $\epsilon$,
at $a_s^2$, $\omega=2$ and so on.  If we generalise this uniform transcendentality to all orders, the highest
power of $\ln(1-z)$ turns out to be $i+j$.    
}
\begin{eqnarray}
\label{PhiBlog}
\Phi_B^c&=& \sum_{i=1}^\infty \hat a_s^i \left({q^2 \over \mu^2}\right)^{i{\epsilon \over 2}}
S_\epsilon^i  \sum_{j=-i}^\infty 
 \sum_{k=0}^{i+j}  \hat \Phi^{c,(i,j)}_k \epsilon^j \ln^{k}(1-z).
 \nonumber\\&&
\end{eqnarray}
Due to this structure,  we find in the successive orders of $a_s$ in $\Delta_c$, there is an increment of two in the power of leading $\ln(1-z)$ terms. 

\textcolor{black}{{\it Multi-scale structure: ---} In \cite{Anastasiou:2014lda} the CFs were computed up to third order in $a_s$ using the method of threshold expansion in dimensional regularisation.  Interestingly, for the diagonal channel,   $\hat \sigma_{gg}$, the results show remarkable structure in terms of $z$ and $\epsilon$.  One finds that  $\hat \sigma_{gg}$ factorises into terms of the form $(1-z)^{\epsilon}$ and functions that depend only on $\epsilon$.  Generalisation to $i$th order in $a_s$ gives factorisation of the form, $\sum_{\eta=2}^{2 i} (1-z)^{\eta \epsilon/2} \chi_i^\eta(\epsilon)$.  The factor $(1-z)^{ \eta \epsilon/2}$ results from soft and collinear configurations of partons at the corresponding soft and collinear scales given by $(q^2 (1-z))^{\eta \epsilon/2}$.  This allows us to 
sum up the $\ln(1-z)$ terms in 
\eqref{PhiBlog} to obtain
\begin{eqnarray}
\label{PhiBomz}
 \Phi^c_B = \sum_{i=1}^\infty \hat a_s^i   \sum_{\eta=2}^{2i} \left( {q^2 (1-z)^{\eta\over i} \over \mu^2}\right)^{{i \epsilon  \over 2}}S_\epsilon^i
 \tilde \varphi_{c,\eta}^{(i)}(\epsilon) 
\end{eqnarray}
The form of the above solution inspired from the structure of fixed order results obtained in \cite{Anastasiou:2014lda}
explicitly reveals the presence of multiple
scales. One finds that every
collinear parton gives $(1-z)^{\epsilon/2}$ and soft parton gives $(1-z)^\epsilon$
while pure virtual contributions to born amplitude give $\delta(1-z)$ and
the hard part from the real emissions gives terms proportional to $(1-z)^\eta, \eta\ge 0$.   At given order $a_s$, we can determine the values of
$\eta$ by counting the  allowed soft and collinear configurations in
that order. The values of $\eta$ extracted from results known to third order can be
used to extrapolate 
to obtain the upper limit on $\eta$ at $i$th order in $a_s$ and it turns out  to be $2 i$.  The coefficients of the scales $\chi_i^\eta(\epsilon)$ can be expanded in powers 
of $\epsilon$.  The singularity structure in $\epsilon$ is completely
determined by the finiteness of mass factorised result.  Note that the
multi-scale structure of the solution is peculiar to the NSV part of the
solution.} 
\textcolor{black}{
{{\it Class of solutions: ---} We observe that the differential equation for ${\cal S}_c$ allows us to construct not just one solution  but a class of solutions.  In the following we construct a set of solutions, called a minimal class,each parametrised by $\alpha$, satisfying the right divergent structure as well as the dependence on $\ln^k(1-z), k=0,1,\cdots$:}
\begin{eqnarray}
\label{alpha}
{\rm{\Phi}}^c_{B,\alpha} = \sum_{i=1}^\infty \hat a_s^i \left( {q^2 (1-z)^\alpha \over \mu^2}\right)^{{i \epsilon  \over 2}} S_{\epsilon}^i
\overline\varphi_{c,\alpha}^{(i)}(z,\epsilon)  \,.
\end{eqnarray}}       
\textcolor{black}{The predictions from these solutions   are found to be independent of choice of $\alpha$ owing to the explicit $z$-dependence of the coefficients $\overline \varphi_{c,\alpha}^{(i)}(z,\epsilon)$ at every order in $\hat a_s$ and in $\epsilon$. 
It is easy to prove that any variation of $\alpha$ in the factor $(1-z)^{i \alpha \epsilon}$ can be 
compensated by suitably adjusting the $z$ independent coefficients of
$\ln(1-z)$ terms in $\overline \varphi_{c,\alpha}^{(i)}(z,\epsilon)$ 
at every order in $\hat a_s$. 
The reason for this is the invariance of the solution under certain  ``gauge like" transformations 
on both $(1-z)^{i \alpha \epsilon}$
and $\overline \varphi_{c,f,\alpha}(z,\epsilon)$ at every order in $\hat a_s$. Because of this invariance, these transformations neither affect the singular nor the finite parts of 
$ {\rm{\Phi}}^c_{B,\alpha}$.  
The NSV part of the solution
given in \eqref{PhiSV1} is a special case where we set $\alpha=2$ which allows us
to have a common factor  $(1-z)^2$ for SV and NSV.}

\textcolor{black}{{\it Integral representation: ---} Having studied the general structure of $\Phi_B^c$, our next task to sum up the series
to obtain a compact integral representation similar to the SV case.  We use $\Phi_B^c$ given in \eqref{PhiSV1} to obtain}
\begin{eqnarray}
\label{phiBint}
  \!\ \mathrm{\Phi}_{B}^{c}
&&=  \int_{\mu_F^2}^{q^2(1-z)^2} \frac{d\lambda^2}{\lambda^2} L^c(a_s(\lambda^2),z)  \nonumber\\&& 
+ \varphi_{f,c}\big(a_s(q^2(1-z)^2),z,\epsilon\big) 
    + \varphi_{s,c}\big(a_s(\mu_F^2),z,\epsilon\big) \,,
\end{eqnarray}
\textcolor{black}{Here, the first two terms are finite as $\epsilon \rightarrow 0$
while $\varphi_{s,c}$ is divergent.  
Since $\mathrm{\Phi}_{B}^{c}$ is RG invariant, $\varphi_{s,c}$ satisfies the RG equation:}
\begin{eqnarray}
\label{RGphis}
\mu_F^2 {d \over d\mu_F^2}  \varphi_{s,c}(a_s(\mu_F^2),z) = L^c (a_s(\mu_F^2),z).
\end{eqnarray}
\textcolor{black}{
Further the  fact that $\Delta_c$ in \eqref{MasterF} is finite at every order in $a_s$
in the limit $\epsilon \rightarrow 0$ allows us
to determine the coefficients $L^c$ in terms of the NSV coefficients $C^c$ and $D^c$ in AP splitting kernels.  We find, at each order in perturbative expansion}
\begin{align}
L^c (a_s(\mu_F^2),z) &= \sum_{i=1}^{\infty}a_s^i(\mu_F^2)L^c_i(z)\nonumber\\
\text{~~with},~~&
L^c_i(z) = C_i^c \ln(1-z) + D_i^c \,,
\end{align}
where the coefficients $C^c_i$ and $D^c_i$ are related to those of cusp $A^c_i$ and collinear $B^c_i$ anomalous dimensions (see \cite{Kodaira:1981nh,Kodaira:1982az,Vogt:2004mw,Ravindran:2004mb,Moch:2005tm,Moch:2004pa,vonManteuffel:2016xki,Das:2020adl} and for beyond three loops, see \cite{vonManteuffel:2020vjv,
Moch:2004pa,Vogt:2004mw,Dokshitzer:2005bf}).
\textcolor{black}{
The finite part $\varphi_{f,c}$ can be expanded in powers of $a_s$:}
\begin{align}
\label{varphiexp}
\varphi_{f,c}(\lambda^2,z) =&
\sum_{i=1}^\infty a_s^i(\lambda^2) \sum_{k=0}^i \varphi_{c,i}^{(k)} \ln^k(1-z) \,,
\end{align}
\textcolor{black}{where the highest power of $\ln(1-z)$ are in accordance with the same in \eqref{PhiBlog}. Defining $\{\mu_i\}=\mu_R,\mu_F$, we get}
\begin{eqnarray}
\label{resumz}
\!\!\!\!\Delta_c(q^2\!,\{\mu_i^2\},z)\!=\!C^c_0(q^2\!,\{\mu_i^2\}) 
{\cal C} \exp \big(2 \Psi^c(q^2\!,\mu_F^2\!,z) \big)\,,
\end{eqnarray}
where
\begin{eqnarray}
\label{phicint}
\Psi^c (q^2,\mu_F^2,z) &=& {1 \over 2}
\int_{\mu_F^2}^{q^2 (1-z)^2} {d \lambda^2 \over \lambda^2} 
	P^{\prime}_{cc} (a_s(\lambda^2),z) 
\nonumber\\&&+ {\cal Q}^c(a_s(q^2 (1-z)^2),z)\,,\nonumber\\
{\rm with} \quad {\cal Q}^c (a_s,z) \!\!
&=& \!\!\ \left(\!{1 \over 1-z}\! \overline G^c_{SV}(a_s)\!\right)_+ \!\!+ \varphi_{f,c}(a_s,z).
\end{eqnarray}
\textcolor{black}{The coefficient $C_0^c$ is $z$ independent coefficient and is expanded in powers of $a_s(\mu_R^2)$ as
$C_0^c(q^2,\mu_R^2,\mu_F^2) = \sum_{i=0}^\infty a_s^i(\mu_R^2) C_{0i}^c(q^2,\mu_R^2,\mu_F^2)$.}
{\color{black} An elaborate discussion on $\rm \Phi^{c}$ can be found in the longer version of the paper \cite{Ajjath:2020ulr}. The integral representation given in \eqref{resumz} is suitable for obtaining certain SV and NSV terms to all orders, which subsequently lead to a framework to resum the diagonal NSV terms \cite{Ajjath:2020ulr}. }

\textcolor{black}{{\it All order predictions: ---}  Given $\Psi^c$ at  order $a_s$, expanding the exponential in powers of $a_s$  we obtain the leading SV terms $({\cal D}_3,{\cal D}_2)$,
$({\cal D}_5,{\cal D}_4),\cdots,({\cal D}_{2i-1},{\cal D}_{2i-2})$ and the leading NSV terms
$\ln^3(1-z),\ln^5(1-z),\cdots,\ln^{2i-1}(1-z)$  at $a_s^2,a_s^3,\cdots,a_s^i$
respectively for all $i$.
Since $C^c_1$ is identically zero, $\ln^{2i}(1-z)$ terms do not contribute for all $i$.
\textcolor{black}{ At this stage, we can ask
whether these predictions will be affected if we include second order result for $\Psi^c$.
Since the power of the leading logarithm at $\varepsilon^{j}$ accuracy
is $2+j$ and hence at $\varepsilon^0$ order the highest logarithm
is $\log^2(1-z)$ we observe that second order result for $\Psi^c$ will only contribute to sub leading logarithms at $a_s^2$, not to leading
ones.  Similarly prediction at third order will also be unaffected
by third order result for $\Psi^c$ and so on.} 
Now from $\Psi^c$ to order $a_s^2$, we can predict
the tower consisting of  $({\cal D}_3$,${\cal D}_2)$, $({\cal D}_5,{\cal D}_4)$,
$\cdots$,$({\cal D}_{2i-3},{\cal D}_{2i-4})$
and of $L^4_z,L^6_z,\cdots,L^{2i-2}_z$ at $a_s^3,a_s^4,\cdots,a_s^i$ respectively for all $i$.\textcolor{black}{Note that even though $L^4_z$ term is absent at the second order in $\Psi^c$ at the accuracy $\varepsilon^0$, we can predict this term simply because of convolutions between ${\cal D}_l$ and $L^m_z$ from first and second order terms in  
$\Psi^c$.}  
Generalising this, if we know $\Psi^c$ up to $n$th order,
we can predict $({\cal D}_{2i-2n+1},{\cal D}_{2i-2n})$ and $L^{2i-n}_z$ at every order in $a_s^i$ for all $i$. 
}

\textcolor{black}{{\it Conclusions ---}In this letter, we have set up a formalism to sum up both SV and NSV logarithms of diagonal CFs of inclusive production of a colorless state in hadron colliders to all orders in perturbative QCD.  The
simple factorisation structure  helped us to set up a Sudakov type
integro-differential equation with respect to $q^2$ for the soft-collinear function.  The later implies a remarkable factorisation of IR singular part in the soft-collinear function to all orders in perturbation theory.   Its solution admits exponential structure and  thanks to uniform transcendentality structure
for the leading logarithms of ${\rm CF}_d$s, we could parametrise the $z$ dependence
of the solution at every order in $a_s$ given an accuracy in $\epsilon$. The resulting integral representation for the solution allows us to predict certain SV and NSV terms to all orders from the knowledge of previous order information and in addition, it will be useful for resummation studies in Mellin-$N$ space.       
Our result will be useful for phenomenological
studies for processes such as Drell-Yan and Higgs boson productions at the LHC.}

\textit {Acknowledgements} --- We  thank  Claude  Duhr  for  useful  discussion  and  his constant help throughout this project.  We thank Claude Duhr and Bernhard Mistlberger for providing third order results for the inclusive reactions.  VR thanks G. Grunberg  for  useful  discussions.   We  would  also like  to  thank  L. Magnea and E. Laenen for their encouragement to work on  this  area.  
\bibliography{NSV_DY_PRL}
\end{document}